\documentclass[a4paper,11pt]{article}

\usepackage[margin=1in]{geometry}

\usepackage[english]{babel}
\usepackage{amssymb}
\usepackage{amsmath}
\usepackage{amsthm}
\usepackage[T1]{fontenc}
\usepackage{ae,aecompl}
\usepackage[colorlinks]{hyperref}
\usepackage{appendix}
\usepackage[superscript,biblabel]{cite}
\usepackage{graphicx}
\usepackage{color}
\usepackage{authblk}
\usepackage{multibib}
\newcites{main}{References}
\newcites{app}{References}

\providecommand\bnabla{\boldsymbol{\nabla}}

\newcommand\ie{i.e.\ }

\newcommand{\pdt}[1]{\ensuremath{\frac{\mbox{$\partial$} #1}{\mbox{$\partial$} t}}}
\newcommand{\pl}{\left(}
\newcommand{\pr}{\right)}
\newcommand{\n}{\nabla}

\newcommand{\vect}[1]{\boldsymbol{#1}}
\newcommand{\vel}{\mathbf{u}}
\newcommand{\velg}{\mathbf{u}_{\perp}}

\newcommand{\vorz}{\zeta}
\newcommand{\moyp}[1]{\overline{#1}}
\newcommand{\moyz}[1]{\left \langle #1 \right \rangle}

\newcommand{\mpeak}{m_{p}}

\newcommand{\lpeak}{\mathcal{L}}
\newcommand{\urms}{\mathcal{U}}
\newcommand{\lrms}{\mathcal{L}}

\newcommand{\Pran}{\mbox{\textit{Pr}}} 
\newcommand{\Ek}{\mbox{\textit{Ek}}}
\newcommand{\Ra}{\mbox{\textit{Ra}}}
\newcommand{\Ro}{\mbox{\textit{Ro}}}

\newcommand{\Rey}{\mbox{\textit{Re}}}
\newcommand{\Bu}{\mbox{\textit{Bu}}}
\newcommand{\Rm}{\mbox{\textit{Rm}}}

\newcommand{\intro}[1]{\textbf{#1}}

\title{Convective Lengthscale in Planetary Cores}
\author[$\ast$,1]{C\'eline Guervilly}
\author[2]{Philippe Cardin}
\author[2]{Nathana\"el Schaeffer}
\affil[1]{School of Mathematics, Statistics and Physics, Newcastle University, Newcastle upon Tyne, NE17RU, UK}
\affil[2]{ISTerre, Universit\'e Grenoble Alpes, CNRS, 38041 Grenoble, France}
\begin{document}

\maketitle

\intro{
Convection is a fundamental physical process in the fluid cores of planets because it is the primary transport mechanism for heat and
chemical species and the primary energy source for planetary magnetic fields.
Key properties of convection, such as the characteristic flow velocity and lengthscale, are poorly quantified in planetary cores due to their strong dependence on
planetary rotation, buoyancy driving and magnetic fields, which are all difficult to model under realistic conditions.
In the absence of strong magnetic fields, the core convective flows are expected to be in a regime of rapidly-rotating turbulence\citemain{Aur15}, which remains largely unexplored to date.
Here we use a combination of numerical models designed to explore this low-viscosity regime to show that the convective lengthscale becomes independent of the viscosity and is entirely determined by the flow velocity and planetary rotation.
For the Earth's core, we find that the characteristic convective lengthscale is approximately 30km
and below this scale, motions are very weak.
The 30-km cut-off scale rules out small-scale dynamo action and supports large-eddy simulations of core dynamics.
Furthermore, it implies that our understanding of magnetic reversals from numerical geodynamo models does not relate to the Earth, because they require too intense flows\citemain{Ols06,Olson2014}.
Our results also indicate that the liquid core of the Moon might still be in an active convective state despite the absence of a present-day dynamo\citemain{Weiss2014}.
}

\begin{figure}
	\centering
	\includegraphics[clip=true,height=11cm]{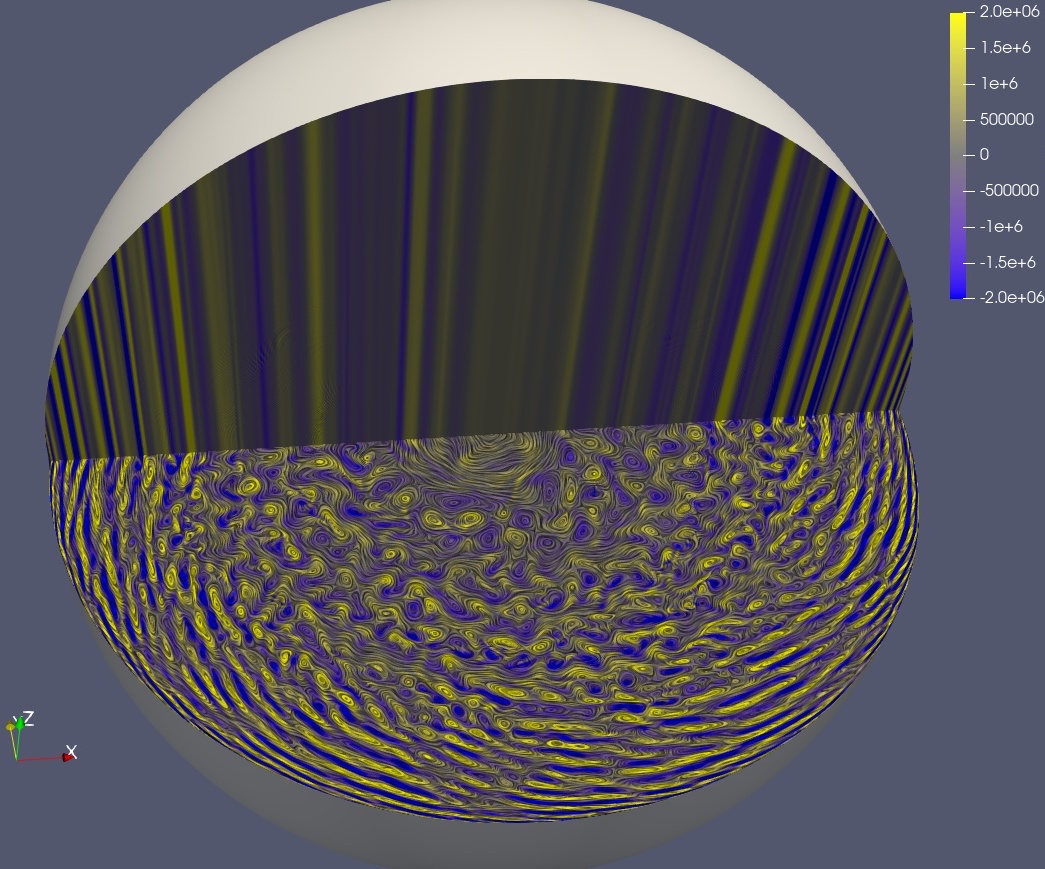}
	\caption{Meridional and equatorial cross-sections of a snapshot of the axial vorticity in the 3D model for $\Ek=10^{-8}$, $\Ra=2\times 10^{10}$, and $\Pran=10^{-2}$. Streamlines have been superimposed in the equatorial plane. The kinetic energy of the velocity projected on a quasi-geostrophic state \mbox{$(\moyz{u_s},\moyz{u_{\phi}},z\beta \moyz{u_s})$} (where the angle brackets denote an axial average) is within $0.2$\% of the total kinetic energy.}
	 \label{fig:view3D}
\end{figure}

Core convection is strongly affected by the rapid planetary rotation through the Proudman-Taylor constraint\citemain{Taylor1922}, 
which obliges the fluid to move in columns with little variation along the rotation axis compared with the orthogonal directions. 
The very low fluid viscosity in liquid cores implies that the convective flows are turbulent, 
but this turbulence differs from both 3D turbulence due to the anisotropy imposed by the rotation and 2D turbulence 
due to the presence of Rossby waves\citemain{Vallis2006}.
Conditions in planetary cores correspond to small Ekman numbers 
(\mbox{$\Ek=\nu/\Omega R^2$} with viscosity $\nu$, rotation rate $\Omega$ and core radius $R$), 
large Reynolds numbers 
(\mbox{$\Rey=\urms R/\nu$} with flow speed $\urms$),
and small Rossby numbers (\mbox{$\Ro=\urms/\Omega R = \Rey\Ek$}), with, for instance, \mbox{$\Ek\approx 10^{-15}$}, \mbox{$\Rey\approx10^9$} and \mbox{$\Ro\approx10^{-6}$} in the Earth's core\citemain{Jon15}.
Numerical models of planetary cores must employ a fluid viscosity that is orders of magnitude larger than realistic values  
to keep the range of time and length scales involved in the dynamics manageable,
with typically \mbox{$\Ek\geq10^{-7}$} and \mbox{$\Rey\leq10^4$}\citemain{Gastine2016}.
Unfortunately this has the undesirable effect that convection properties are still controlled by the viscosity\citemain{King2013b,Oruba2014}.
To go beyond this range of parameters and into the rapidly-rotating turbulent convection regime, in which the fluid viscosity plays a sub-dominant role,
we use a combination of a state-of-the-art 3D model\citemain{Kaplan2017} down to $\Ek=10^{-8}$ supplemented by
a simplified model of rotating convection\citemain{Or87} down to $\Ek=10^{-11}$ using a quasi-geostrophic (QG) approximation. Here the QG approximation
means that the axial vorticity is invariant along the rotation axis in accordance with the Proudman-Taylor constraint.
This approximation is well supported by the results of the 3D model shown in Figure~\ref{fig:view3D}.

\begin{figure}
	\centering
	\includegraphics[clip=true,height=11cm]{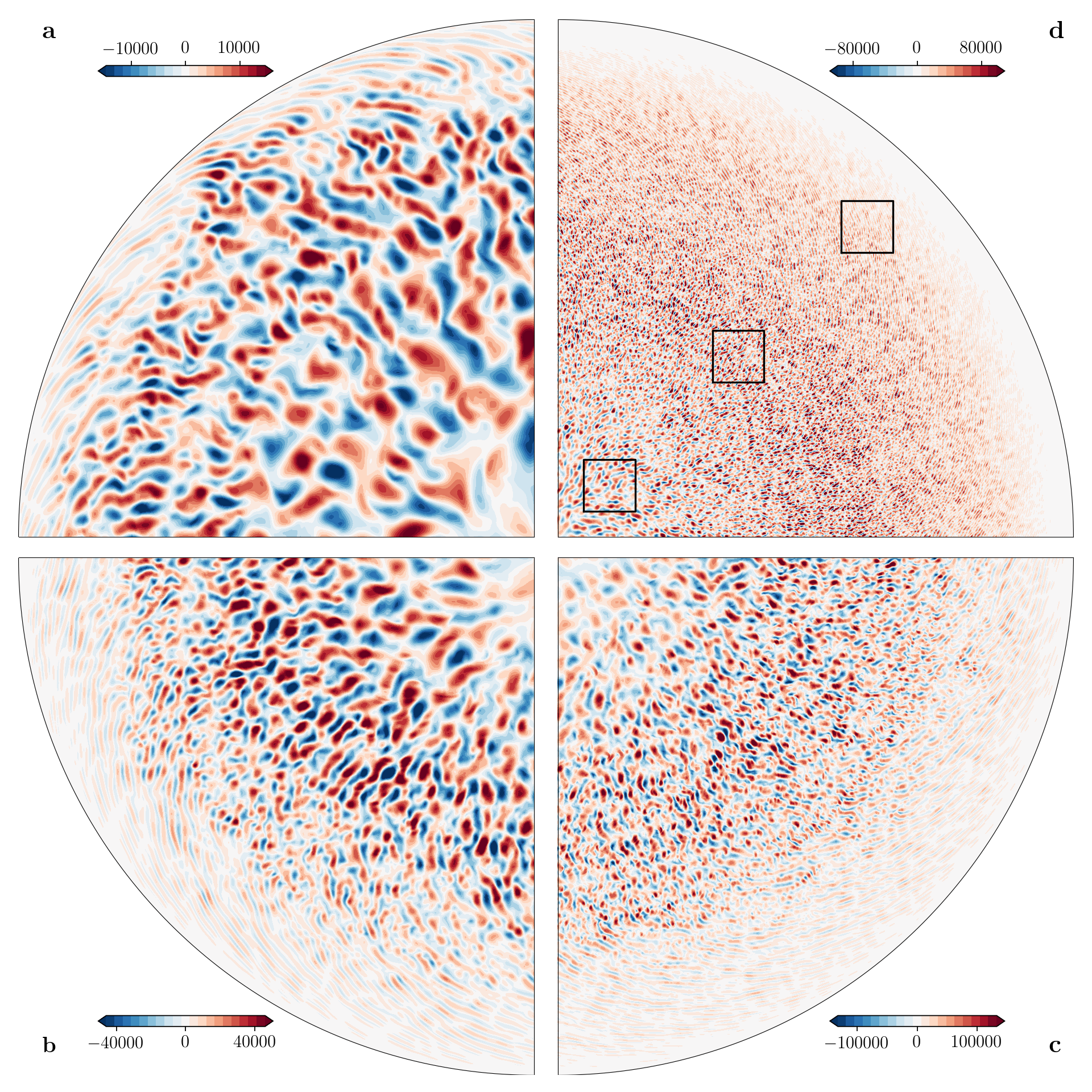}
	\hspace{1cm}
	\includegraphics[clip=true,height=11cm]{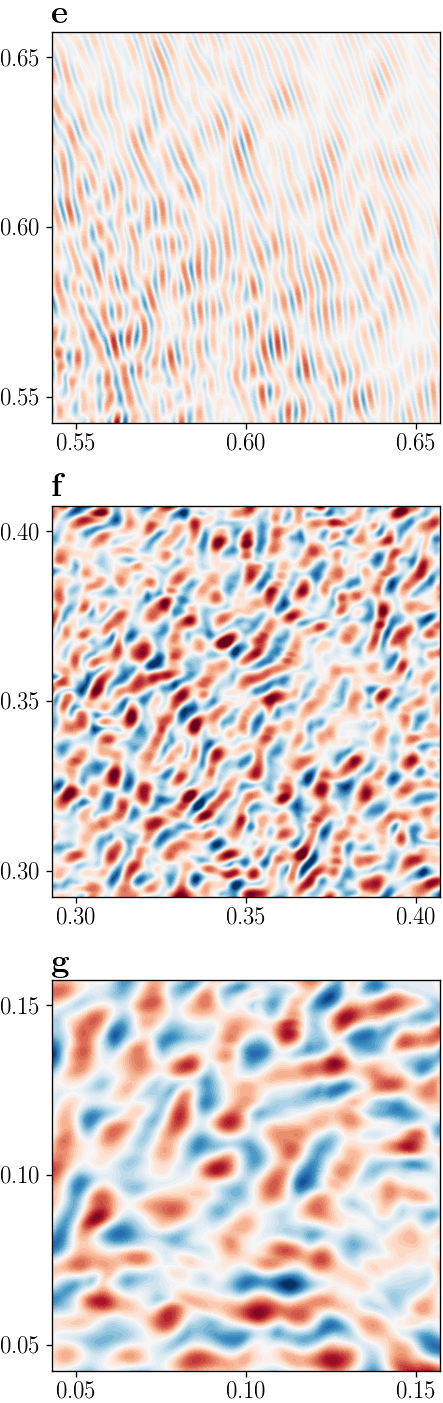}
	\caption{Snapshots of the radial velocity in a quarter of the equatorial plane during the statistically steady phase for 
	a) $\Ek=10^{-8}$, $\Ra=2.5\times10^{10}$ (3D model),
	b) $\Ek=10^{-9}$, $\Ra=2.7\times10^{11}$ (QG model),
	c) $\Ek=10^{-10}$, $\Ra=6.3\times10^{12}$ (QG model) and
	d) $\Ek=10^{-11}$, $\Ra=5.25\times10^{13}$ (QG model).
	Close-ups of the equatorial plane are shown in e-g for the same parameters as in d; e shows the outer conduction-dominated region where the dynamics is dominated by Rossby waves,
	and f-g the inner convective region. 
	The Prandtl number is $10^{-2}$ in all cases.
	The colorbars give the radial velocity normalised by the viscous velocity scale (\ie corresponding to a Reynolds number).}
	 \label{fig:ur}
\end{figure}

\begin{figure}
	\centering
	\includegraphics[clip=true,width=0.5\textwidth]{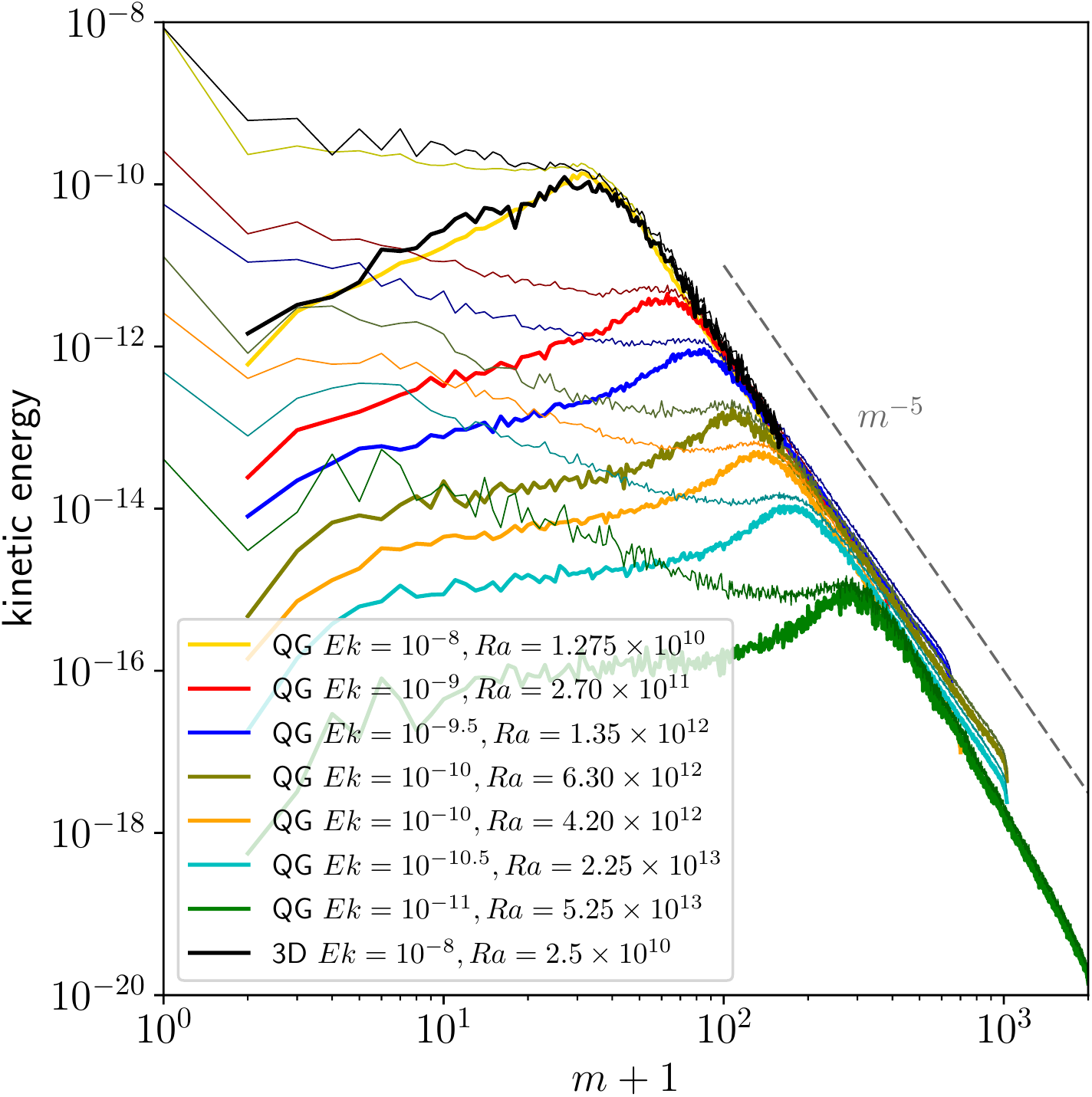}
	\caption{Power spectra of the total kinetic energy (thin line) and radial kinetic energy (thick line) at $s=0.5$
	as a function of the azimuthal wavenumber $m$ at $s=0.5$ for simulations with different Ekman and Rayleigh numbers for $\Pran=10^{-2}$ performed with the 3D and QG models.
	The kinetic energy is averaged in time and normalised by $\rho(\Omega R)^2/2$.}
	 \label{fig:spec}
\end{figure}

For the low Ekman numbers studied here, convection is always in a turbulent state, even near onset\citemain{Guervilly2016,Kaplan2017}, and $\Rey\geq10^3$.
The convection takes the form of vortical plumes that are radially elongated on scales much shorter than the outer radius (Figure~\ref{fig:ur}). 
At large radius, the steepening of boundary slope leads to rapid changes in the column height, which inhibit convection\citemain{Jon15}. 
The dynamics there mainly consists of Rossby waves, which appear as elongated vortices with a prograde tilt (Figure~\ref{fig:ur}e).
Their radial velocity is relatively small so conduction dominates the heat transport in the outer part of the equatorial plane \citemain{Guervilly2017}. 
Hereafter we solely consider the dynamics of the inner convective region, which grows wider with increasing Rayleigh number ($\Ra$, which measures the strength of the buoyancy
driving with respect to dissipative effects). 
The azimuthal lengthscale of the convective flows decreases notably with radius (Figures~\ref{fig:ur}f-g)
to minimise the changes in the column height.
At lower $\Ek$, the scale of the convective flow is visibly smaller. We find that the convective lengthscale is controlled by 
the Rossby number, rather than by any viscous effect.
The flows shown in Figures~\ref{fig:view3D} and \ref{fig:ur} are snapshots taken once the system has reached a statistically steady state, and are 
entirely unlike the linear viscous mode at the convection onset, 
which consists of drifting columns with a narrow azimuthal lengthscale that scales as \mbox{$\Ek^{1/3}$}\citemain{Zha92,Jon00}.
The convective lengthscale increases with the buoyancy driving 
as seen on the power spectra of the total and radial kinetic energies in Figure~\ref{fig:spec}.
The peak of the radial kinetic energy moves to smaller wavenumber for increasing $\Ra$, as can be observed
for the two different Rayleigh numbers shown at $\Ek=10^{-10}$, and is located at significantly smaller wavenumber ($m=133$ and $106$ for the smaller and larger $\Ra$) 
than the wavenumber of the marginal linear viscous mode at onset ($m_c=258$). 
Remarkably, the spectra at different $\Ek$ and $\Ra$ superpose well at wavenumbers larger than the peak,
and follow a steep slope $m^{-5}$\citemain{Sch05b}.
There is therefore a well-defined characteristic convective lengthscale
that carries most of the radial kinetic energy, and below this scale,
the velocity amplitude drops very rapidly.
This characteristic lengthscale
is thus a limit below which only weak convective motions occur,
thereby restricting viscous dissipation in the bulk.
At wavenumbers smaller than the peak, the velocity becomes anisotropic with a dominant azimuthal component.
The kinetic energy is transferred to larger scales, 
where the dynamics is dominated by propagating Rossby waves, and viscous dissipation can occur in the boundary layers.

\begin{figure}
	\centering
	\includegraphics[clip=true,width=0.49\textwidth]{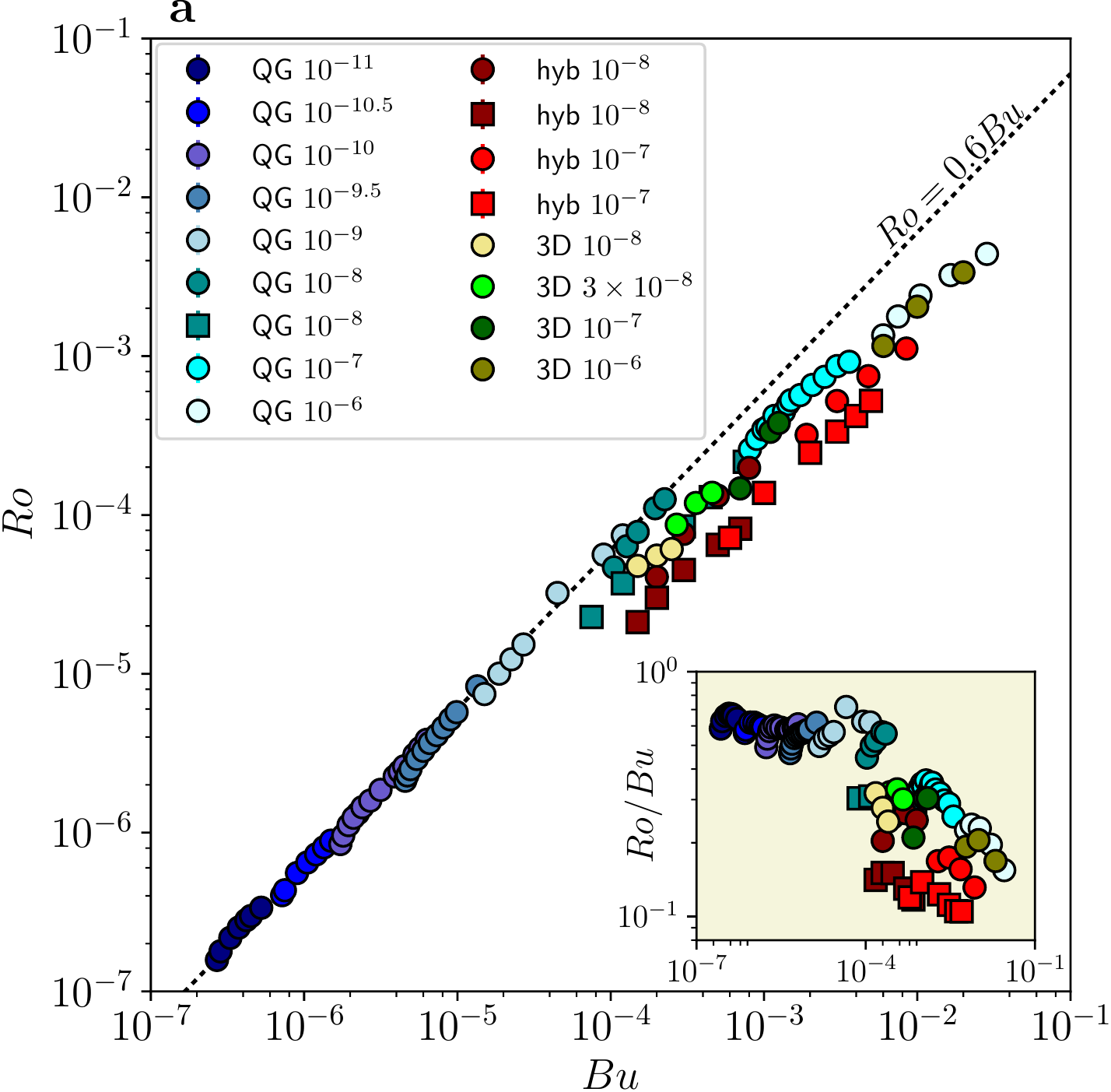}
	\includegraphics[clip=true,width=0.49\textwidth]{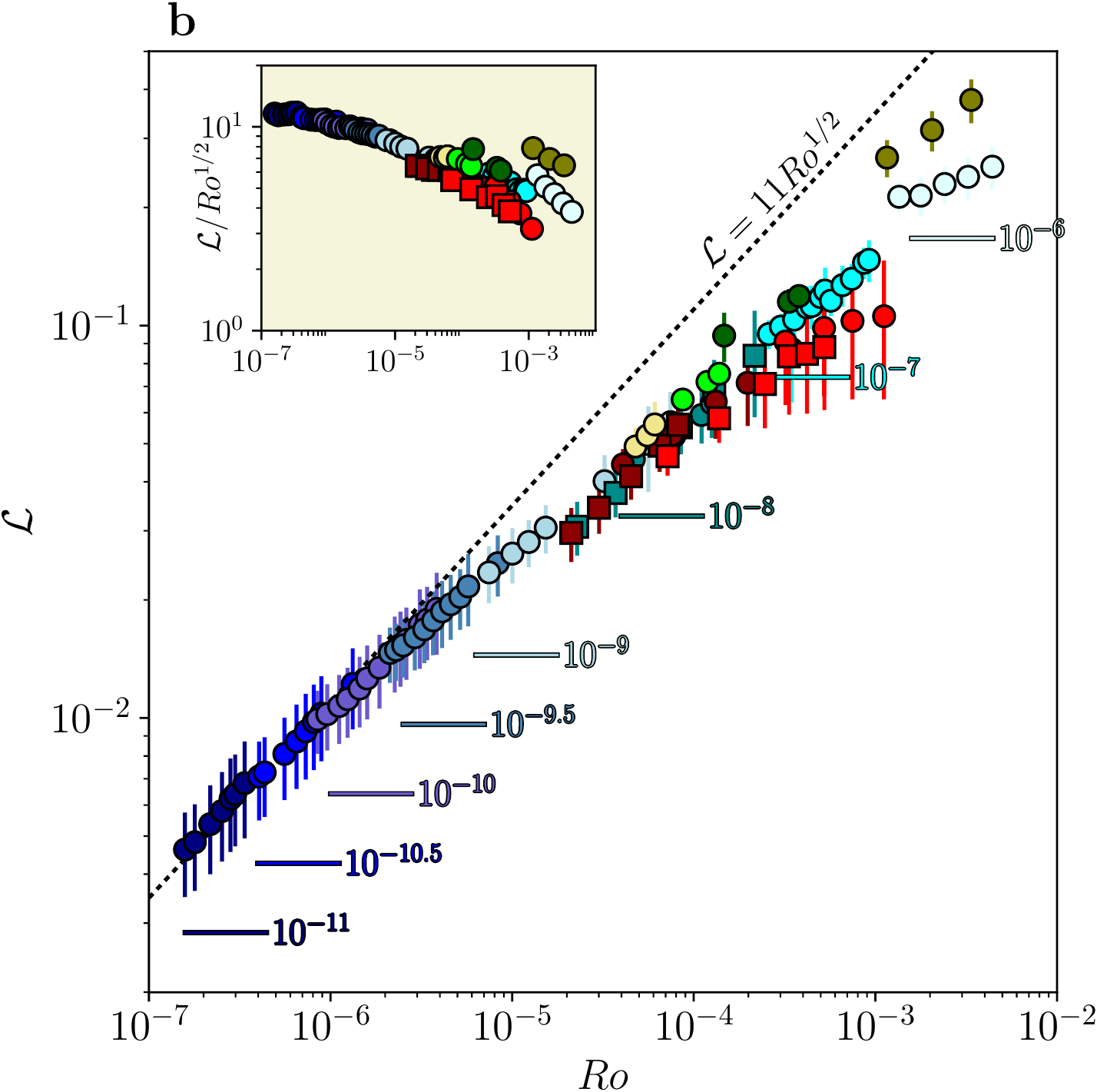}
	\caption{a) Rossby number as a function of the buoyancy parameter
	and b) characteristic convective lengthscale as a function of the Rossby number
	in simulations performed with the 3D model (green points) for $\Ek\in[10^{-8},10^{-6}]$, the QG model (blue) for $\Ek\in[10^{-11},10^{-6}]$, 
	and the hybrid model (red) for $\Ek\in[10^{-8},10^{-7}]$. 
	Marker colours correspond to Ekman numbers (values given in the legend) and marker shapes to Prandtl numbers (circles: $\Pran=10^{-2}$
	and squares: $\Pran=10^{-1}$).
	In b, the convective scale is averaged in radius between $s=0.1$ and $0.6$ and the vertical bars give the standard deviation in this interval. 
	The horizontal lines give the lengthscale corresponding to the marginal viscous linear mode at onset at $s=0.5$ for the Ekman number indicated on the right and $\Pran=10^{-2}$.
	The insets show the data compensated by the theoretical scaling as a function of a) $\Bu$ and b) $\Ro$.}
	 \label{fig:scaling}
\end{figure}

In the rapidly-rotating turbulent regime, the increase of the convective lengthscale with the buoyancy driving is expected 
from scaling arguments\citemain{Stevenson1979,Ing82,Aub01},
which assume that the production of axial vorticity is governed by a triple balance 
between the vorticity advection, vortex stretching and vorticity generation by buoyancy, and is thus independent of viscosity.
The scaling gives a convective lengthscale that depends on the flow velocity
as \mbox{$\lrms \sim \pl \Ro/|\beta| \pr^{1/2}$}, where $\beta$ is a geometrical parameter related to the boundary slope (see Methods).
This lengthscale 
is consistent with the $-5$ slope observed on the power spectra of the kinetic energy.
Assuming that the transport in the fluid bulk controls the heat transfer\citemain{Julien2012}, 
the scaling uses a balance between the nonlinear advection of temperature and the 
transport of the mean temperature background to obtain \mbox{$\Rey \sim \Ra \Ek/\Pran$}, or simply 
\mbox{$\Ro \sim \Bu$} with the viscosity-free buoyancy parameter \mbox{$\Bu=\Ra \Ek^2/\Pran$}.
The theoretical scaling law is tested in figure~\ref{fig:scaling}
against results obtained with the 3D and QG models and published results obtained with a hybrid model that uses the QG approximation coupled to
the 3D temperature\citemain{Guervilly2017}. 
The characteristic convective lengthscale corresponds to the peak of the radial kinetic energy spectra.
Points obtained at different $\Ek$ collapse well into a single curve, especially for \mbox{$\Ek<10^{-9}$}, showing that the dependence of 
the results on the viscosity becomes negligible when core conditions are approached. 
Importantly, the good agreement obtained between the different numerical models where the parameters overlap supports the use of the QG approximation 
for modelling rapidly-rotating convection. 
The data of the velocity and lengthscale compensated by their respective theoretical scaling laws align on a plateau at small $\Ek$, and thus 
indicate that the agreement between the simulations and the theoretical scaling improves progressively as $\Ek$ decreases.
The dependence on the Prandtl number can be assessed to a small extent as $\Pran$ is varied by one decade at most:
the lengthscale does not show a significant dependence on $\Pran$, but 
for the velocity, the points all follow a similar slope but cases with larger $\Pran$ tend to have a smaller prefactor.
To avoid the ``shingling" effect that occurs when using diffusion-free parameters \citemain{Cheng2016}, the scaling of the Reynolds number is
shown in Extended Data Figure~\ref{extfig:scalingRe} and confirms the overlap of the data for \mbox{$\Ek\leq 10^{-9}$} and the good agreement with the
exponent predicted by the theoretical scaling.
While the lengthscale measurement corresponds to an azimuthal size, we confirm that 
the radial lengthscale obtained from radial correlations is in good agreement with this azimuthal scale in Extended Data Figure~\ref{extfig:radialscale}.
The radial dependence of the lengthscale observed on the equatorial maps of Figure~\ref{fig:ur}
is also in agreement with the theoretical dependence of the lengthscale on \mbox{$|\beta|^{-1/2}$} as shown in the Extended Data Figure~\ref{extfig:LRo}.

While current 3D models cannot compute values of $\Ek$ below $10^{-8}$, 
our simplified numerical model allows us to reach $\Ek=10^{-11}$, which is approximately the value for the core of the Moon\citemain{Weber2011}. 
The lack of a present-day lunar dynamo places an upper limit on the flow speed in the liquid core of the Moon. 
Assuming that the minimum magnetic Reynolds number (\mbox{$\Rm=\urms R/\eta$} with $\eta$ the magnetic diffusivity) 
required for dynamo action driven by convection is $10$\citemain{Chr06}, the upper bound for $\Ro$ in the Moon core is 
$10^{-4}$. 
According to our results, the convective structures are thus expected 
to be smaller than $0.1R\approx10$km. 
From the scaling of the flow speed with the input parameters (see Extended Data Figure~\ref{extfig:scalingRe}), we can also deduce
that the Rayleigh number must be smaller than about $10^{16}$ in the Moon core (using \mbox{$\Pran=10^{-2}$}). This is approximately a hundred times larger than 
the critical value at the linear convection onset, so vigorous convection is possible in the Moon liquid core, although not sufficiently vigorous to produce a magnetic field.


The simple dependence of the convective lengthscale on $\Ro$ places new constrains on the flows within the Earth's core. Characteristic flow speeds at the core-mantle boundary inferred from observations of the secular variation of the geomagnetic field have $\Ro\approx10^{-6}$\citemain{Holme2006}. 
This value corresponds to a characteristic convective
lengthscale of approximately \mbox{$0.01R\approx30$km}, at which the local magnetic Reynolds number is approximately 10\citemain{Olson15}.
This result has three important implications for our understanding of the geodynamo.
First, the local Rossby number associated with this scale,
\mbox{$\Ro_\ell = \Ro R/\lrms \approx 10^{-4}$}, is considerably smaller
than the value $\Ro_\ell = 0.1$ required for polarity reversals of the magnetic field in viscously-controlled geodynamo simulations\citemain{Ols06,Olson2014}.
This discrepancy highlights the need for further geodynamo simulations in the rapidly-rotating turbulent regime to model Earth-like polarity reversals.
Second, while information about core flows is spatially limited due to 
the 1000-km resolution limit of geomagnetic flux patches at the core-mantle boundary, the 30-km convective lengthscale is a lower limit 
for the energy-carrying lengthscales.
This lower limit provides a useful estimate for the viscous dissipation in the core\citemain{Nat15}.
The viscous dissipation in the bulk for $\lrms \approx30$~km is of the order of 1~kW, much lower than the 1~MW dissipated within the laminar Ekman layers.
Third, in the presence of magnetic fields, the convective lengthscale is expected to increase \citemain{Cha61,Yad16b}, so the 30-km scale will likely remain a lower limit.
The steepness of the kinetic energy spectrum beyond this scale implies that no small-scale dynamo can operate for such rapidly-rotating turbulent convection.
Finally, this relatively large convective lengthscale (\ie much larger than the viscous scale) supports the use of small-scale parameterisation such as those adopted in recent simulations\citemain{Kaplan2017,Aubert2017}.

\bibliographystylemain{unsrt}
\bibliographymain{ref}

\section*{Acknowledgements}
C.G. was supported by the Natural Environment Research Council under grant NE/M017893/1.
P.C. and N.S. were supported by the French {\it Agence Nationale de la Recherche} under grants ANR-13-BS06-0010 (TuDy) and ANR-14-CE33-0012 (MagLune).
N.S. acknowledges GENCI for access to resource Occigen (CINES) under grant A0020407382 and A0040407382.
This research made use of the Rocket High Performance Computing service at Newcastle University, the ARCHER UK National Supercomputing Service (\url{http://www.archer.ac.uk}), and the DiRAC Data Centric system at Durham University, operated by the Institute for Computational Cosmology on behalf of the STFC DiRAC HPC Facility (\url{www.dirac.ac.uk}).
This equipment was funded by BIS National E-infrastructure capital grant ST/K00042X/1, STFC capital grant ST/H008519/1, and STFC DiRAC Operations grant ST/K003267/1 and Durham University. DiRAC is part of the National E-Infrastructure.
Parts of the computations were also performed on the Froggy platform of CIMENT (\url{https://ciment.ujf-grenoble.fr}), supported by the Rh\^ one-Alpes region (CPER07\_13 CIRA), OSUG@2020 LabEx (ANR10 LABX56) and Equip@Meso (ANR10 EQPX-29-01).
ISTerre is part of Labex OSUG@2020 (ANR10 LABX56).

\section*{Author contributions}
C.G. and P.C. performed the numerical simulations with the QG code. N.S. performed the numerical simulations with the 3D code. The manuscript was mainly written by C.G.
All authors contributed to the analysis of the data and the preparation of the manuscript.

\section*{Competing financial interests}
The authors declare no competing financial interests.

\section*{Additional information}
Correspondence and requests for materials should be addressed to C.G. (celine.guervilly@ncl.ac.uk).

\newpage

\section*{Method}

We model Boussinesq convection driven by homogeneous internal heating
in a full sphere geometry. This problem is relevant for planetary cores without a solid inner core,
and thus, for most of the Earth's history \citeapp{Labrosse2015}. 
The sphere rotates at the rate $\Omega$ around the axis directed along $\hat{e}_z$.
The acceleration due to gravity is radial and increases linearly, $\vect{g} = - g_0 r\hat{e}_r$.
The governing equations are written in dimensionless form, obtained by scaling lengths
with the outer radius $R$, times with $R^2/\nu$ where $\nu$ is the 
fluid kinematic viscosity, and temperature with \mbox{$\nu S R^2/(6\rho C_p\kappa^2)$},
where $S$ is the internal volumetric heating, $\kappa$ the thermal diffusivity, $\rho$ the density, 
and $C_p$ the heat capacity at constant pressure. 
The dimensionless numbers are: the Ekman number, \mbox{$\Ek=\nu/(\Omega R^2)$},
the Rayleigh number, \mbox{$\Ra=\alpha  g_0 S R^6/(6 \rho C_p \nu \kappa^2$)}, where 
$\alpha$ is the thermal expansion coefficient, and the Prandtl number, \mbox{$\Pran=\nu/\kappa$}.
This study focuses on Prandtl numbers smaller that unity, which are relevant for thermal convection of liquid metal cores\citeapp{Poz12}.
The system of dimensionless equations is:  
\begin{eqnarray}
	&&\pdt{\vel} + \pl \vel \cdot \boldsymbol{\n} \pr \vel + \frac{2}{\Ek}\hat{e}_z \times \vel 
	= - \boldsymbol{\n} p + \boldsymbol{\n^2} \vel + \Ra \Theta \mathbf{r} ,
	\label{eq:NS}
	\\
	&& \nabla \cdot \vel = 0,
	\\
	&& \pdt{\Theta} + \vel \cdot \nabla \Theta -\frac{2}{\Pran} r u_r = \frac{1}{\Pran} \nabla^2 \Theta,
	\label{eq:T3D}
\end{eqnarray}
where $\vel$ is the velocity field, $p$ the  pressure, and
$\Theta$ the temperature perturbation relative to the static temperature $T_s = (1-r^2)/\Pran$. 
We use no-slip boundary conditions and fixed temperature at the outer boundary.

For the 3D simulations, we use the code XSHELLS\citemain{Kaplan2017}, which solves Equations~\eqref{eq:NS}-\eqref{eq:T3D} using
finite differences in the radial direction and spherical harmonic expansion\citeapp{Sch13}.
The input parameters and numerical resolutions used for the 3D simulations are given in the Extended Data Table~\ref{tab:3D}.
In the 3D simulations, the Prandtl number is fixed to $\Pran=10^{-2}$ and the Ekman number is varied between $10^{-6}$ and $10^{-8}$.
To speed-up these 3D simulations, we increase the viscosity of the smallest scales in the last 10\% of the spectrum\citemain{Kaplan2017} (see Extended Data Table~\ref{tab:3D}).

For simulations at smaller Ekman numbers, we assume that the rotational constraint is such that
the variations of the velocity along the axial direction are small compared with 
the variations along the orthogonal directions.
We use the quasi-geostrophic (QG) approximation for rapidly-rotating spherical convection 
developed from the Busse \citeapp{Bus70} annulus model \citemain{Or87}\citeapp{Car94}, and 
widely used in the context of planetary core convection \citeapp{Aub03,Mor04,Pla08,Gil06,Cal12}.
The dynamics is assumed to be dominated by the geostrophic balance, \ie the
Coriolis force balances the pressure gradient at leading order. The leading-order velocity $\velg$
is invariant along $z$ and $\velg = (u_s,u_{\phi},0)$ in cylindrical polar coordinates.
Quasi-geostrophic convection is driven by the cylindrical component of gravity, \mbox{$-g_0 s$}.
By taking the axial average of the $z$-component of the curl of the Navier-Stokes equation,
we obtain the equation for the leading-order axial vorticity, 
$\vorz$,
\begin{equation}
	\pdt{\vorz} + \pl \velg \cdot \bnabla_{\perp} \pr \vorz 
	- \pl \frac{2}{\Ek} + \vorz \pr \moyz{\frac{\partial u_z}{\partial z}}
	= \nabla_{\perp}^2 \vorz - \Ra \moyz{\frac{\partial \Theta}{\partial \phi}},
	\label{eq:vorz}
\end{equation}
where  $\bnabla_{\perp} f \equiv (\partial_s f, \partial_{\phi} f/s,0 )$,
$\nabla^2_{\perp} f \equiv \partial^2_s f + s^{-1} \partial_s f + s^{-2} \partial^2_{\phi} f$,
and the angle brackets denote an axial average between $\pm H$ with $H=\sqrt{1-s^2}$ the 
height of the spherical boundary from the equatorial plane.

The velocity is described by a streamfunction $\psi$ that models the 
non-axisymmetric (\ie $\phi$-dependent) components with the addition of an axisymmetric azimuthal flow,
$\moyp{u_{\phi}}$, where the overbar denotes an azimuthal average,
\begin{equation}
	\velg = \frac{1}{H} \bnabla \times \pl H \psi \hat{e}_z \pr + \moyp{u_{\phi}} \vect{e}_{\phi}.
	\label{eq:defPsi}
\end{equation}
This choice of the streamfunction accounts for mass conservation at the outer boundary \citeapp{Sch05}.
We assume that the axial velocity $u_z$ is linear in $z$ and has two contributions: 
the main contribution comes from mass conservation at the outer boundary and is proportional to $\beta=H'/H$; 
the second contribution is due to Ekman pumping, which is produced by the viscous boundary layer and scales as $\Ek^{1/2}$.
The Ekman pumping is parametrised by the formula obtained by asymptotic methods in the limit of small $\Ek$
for a linear Ekman layer \citeapp{Gre68}.

The streamfunction $\psi$ only describes the non-axisymmetric motions, so the axisymmetric 
azimuthal velocity, $\moyp{u_{\phi}}$, is obtained by
taking the azimuthal and axial averages of the $\phi$-component of the Navier-Stokes equation to give
\begin{equation}
 \pdt{\moyp{u_{\phi}}}
 +   \moyp{u_{s} \frac{\partial u_{\phi}}{\partial s}} + \moyp{\frac{u_{s} u_{\phi}}{s}}
 = \n^2 \moyp{u_{\phi}} - \frac{\moyp{u_{\phi}}}{s^2} - \frac{1}{\Ek^{1/2} H^{3/2}} \moyp{u_{\phi}},
\label{eq:uzonal}
\end{equation}
where the last term on the right-hand side corresponds to the Coriolis term simplified using 
mass conservation \citeapp{Aub03}.

The equation for the  temperature perturbation $\Theta$ in the quasi-geostrophic model is obtained by taking the axial average of the 
temperature equation and assuming that $\Theta$ is invariant along $z$ to obtain 
\begin{equation}
 \pdt{\Theta}+ \vel \cdot \bnabla_{\perp} \Theta - \frac{4}{3\Pran} s u_s = 
	\frac{1}{\Pran}\n_{\perp}^2 \Theta.
	\label{eq:T}
\end{equation}
Note that here we use the gradient of the $z$-averaged static temperature profile, \mbox{$\moyz{T_s}'=-4s/(3\Pran)$}, rather than the gradient of the $z$-invariant static temperature profile, \mbox{$(T^{2d}_s)' = -3s/\Pran$}, to allow for a direct comparison of the Rayleigh numbers used in the different models.
The assumption that $\Theta$ is invariant along $z$ is not rigorously justified and is used
for numerical convenience because it permits us to treat the numerical problem in two dimensions, thereby 
considerably reducing the computational load. 
The evolution equation for the streamfunction, axisymmetric velocity and temperature are solved 
on a 2D grid in the equatorial plane. 
The QG code uses a pseudo-spectral code with a Fourier decomposition in the azimuthal direction and
a second-order finite-difference scheme in radius with irregular spacing.
The input parameters and the numerical resolutions used for the QG simulations are given in the Extended Data Table~\ref{tab:QG}.
The Prandtl number is varied between $\Pran=10^{-1}$ and $10^{-2}$ and the Ekman number is varied between $10^{-6}$ and $10^{-11}$,
allowing an overlap with the 3D simulations over 2 decades in $\Ek$.

The influence of the assumption of $z$-invariance of $\Theta$ on the QG results is tested by comparing   
the QG results with published results obtained with a hybrid QG-3D model\citemain{Guervilly2017} at $\Ek\in[10^{-8},10^{-7}]$.
In the hybrid model, the temperature is solved in 3D and coupled to the QG implementation for the velocity.
Figure~\ref{fig:scaling} shows good agreement obtained between the QG and hybrid results for overlapping parameters,
demonstrating that, while this assumption is not mathematically justified, it does not significantly influence QG convection.


The output parameters are given in Extended Data Tables~\ref{tab:QG} and~\ref{tab:3D}.
The characteristic velocity $\urms$ used to calculate the Rossby and Reynolds numbers
is based on the r.m.s.\,radial velocity averaged in volume and time over at least 10 convective turnover timescales. 
The convective lengthscale is calculated as $\lpeak(s)=\pi s/\mpeak(s)$, where $\mpeak$ is the 
wavenumber at the peak of the radial kinetic energy spectrum. The peak is determined by smoothing the radial kinetic energy spectra
with a polynomial of degree $14$. 

The radial lengthscale of the convective flow $\lpeak_r(s)$ is calculated using the auto-correlation function $f$ of the radial component of the velocity field. For a given radius $s$, we calculate 
\begin{equation}
	f(ds) = \moyp{u_s(s,\phi,t) u_s(s+ds,\phi,t)},
\end{equation}
where the overbar denotes an azimuthal average. Snapshots covering at least 2 dynamical timescales are used to compute the temporal average. 
$\lpeak_r(s)$ is the full width at half maximum of $f$.

We estimate the bulk viscous dissipation at the lengthscale $\ell$ as $D_{visc} = \rho \nu U^2 V/\ell^2$ and the dissipation in the laminar Ekman layers as $D_{Ek} = \rho (\nu \Omega)^{1/2} U^2 A$, where $V$ is the volume of the core, $A$ the surface area at the core-mantle boundary and $U \approx 20$~km/yr a typical velocity of the core flow\citemain{Holme2006}.

\section*{Data availability}
Synthetic data is provided in the Extended Data Tables.
Any additional data supporting the findings of this study are available from the corresponding author on request.
The 3D numerical code XSHELLS is freely available at \url{https://bitbucket.org/nschaeff/xshells}.

\bibliographystyleapp{unsrt}
\bibliographyapp{ref}

\newpage
\setcounter{figure}{0}

\begin{figure}[h]
	\centering
	\includegraphics[clip=true,width=0.6\textwidth]{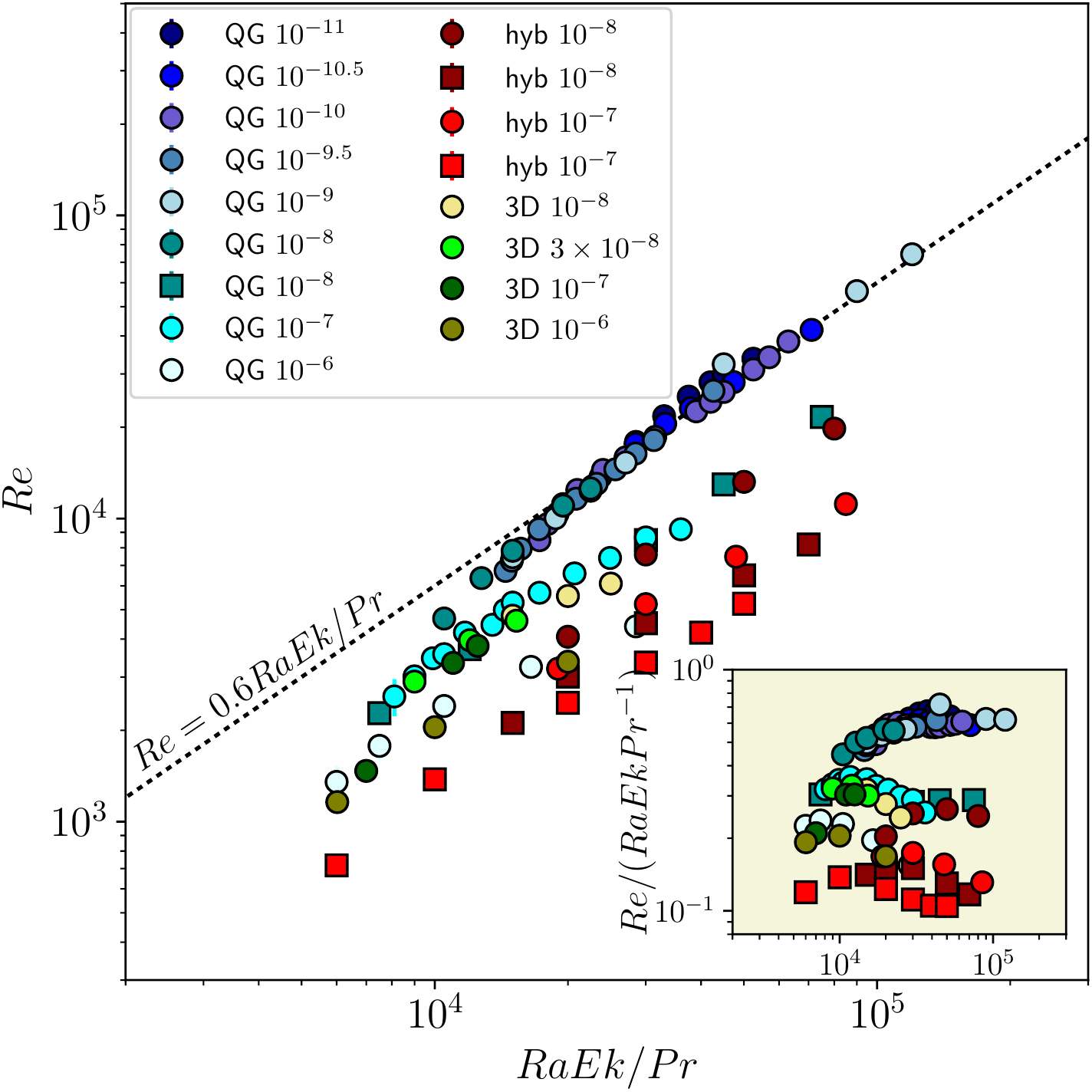}
	\caption{Reynolds number as a function of $\Ra\Ek/\Pran$
	in simulations performed with the 3D model (green points) for $\Ek\in[10^{-8},10^{-6}]$,
	the QG model (blue) for $\Ek\in[10^{-11},10^{-6}]$, and the hybrid model (red) for $\Ek\in[10^{-8},10^{-7}]$. 
	Marker colours correspond to Ekman numbers (values given in the legend) and marker shapes to Prandtl numbers (circles: $\Pran=10^{-2}$
	and squares: $\Pran=10^{-1}$).
	The inset shows the data compensated by the theoretical scaling as a function of $\Ra\Ek/\Pran$.}
	 \label{extfig:scalingRe}
\end{figure}

\section*{Extended Data Figure~\ref{extfig:scalingRe}: Scaling of the Reynolds number}

\begin{figure}[h]
	\centering
	\includegraphics[clip=true,width=0.6\textwidth]{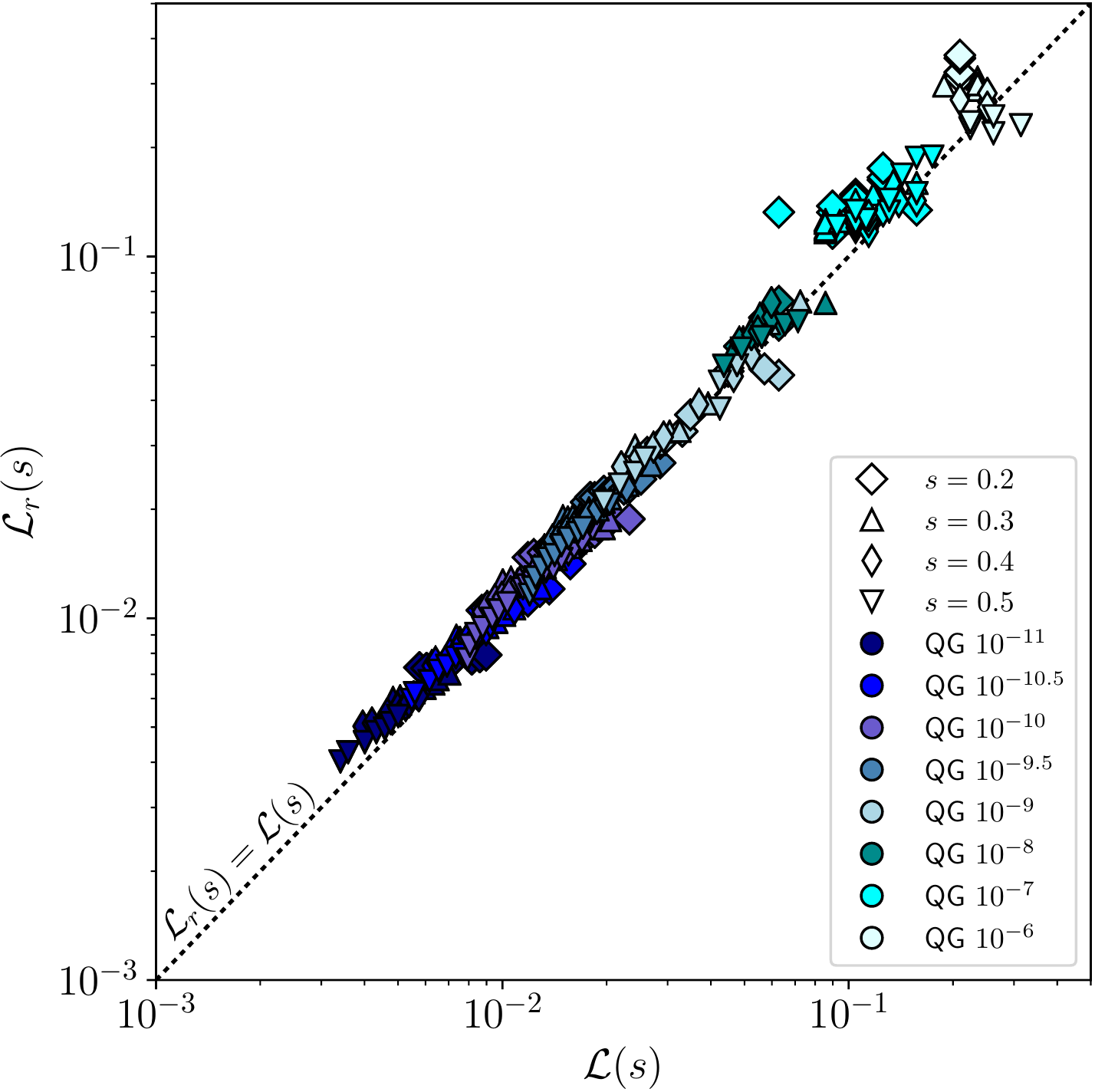}
	\caption{Radial scale of the convective flows $\lpeak_r(s)$ as a function of the azimuthal lengthscale $\lpeak(s)$ obtained with the QG model at different radii.
	Marker colours correspond to Ekman numbers (with $\Pran=10^{-2}$) and marker shapes to different radii.
	The radial scale is calculated from auto-correlation functions of the radial velocity. 
	The convective lengthscale corresponds to an azimuthal scale calculated from the peak of the power spectra of the radial kinetic energy 
	at the radius $s$.}
	 \label{extfig:radialscale}
\end{figure}

\newpage
\section*{Extended Data Figure~\ref{extfig:radialscale}: Radial lengthscale determined by radial correlation}

\begin{figure}[h]
	\centering
	\includegraphics[clip=true,width=0.6\textwidth]{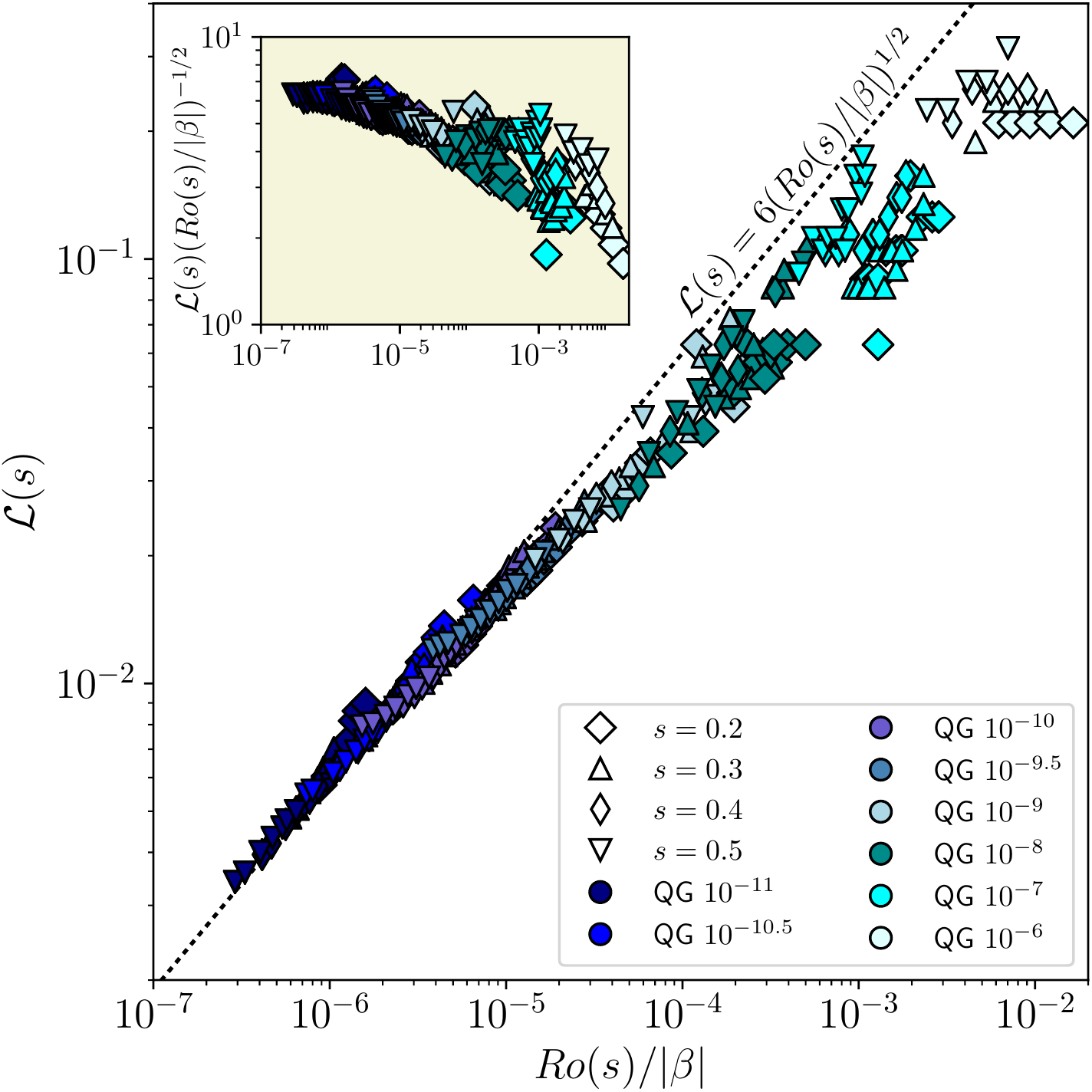}
	\caption{Convective lengthscale $\lpeak(s)$ as a function of $\Ro(s)/|\beta|$ obtained with the QG model 
	at different radii. 
	Marker colours correspond to Ekman numbers (with $\Pran=10^{-1}$ or $10^{-2}$) and marker shapes to different radii. 
	The inset shows the lengthscale compensated by the theoretical scaling as function of $\Ro(s)/|\beta|$.
	The convective lengthscale corresponds to an azimuthal scale calculated from the peak of the power spectra of the radial kinetic energy 
	at the radius $s$.}
	 \label{extfig:LRo}
\end{figure}

\newpage
\section*{Extended Data Figure~\ref{extfig:LRo}: Variation of the convective lengthscale with radius}

\newpage
\section*{Extended Data Table~\ref{tab:QG}: Details of the QG numerical simulations}

\begin{table}[!ht]
\begin{center}
\begin{tabular}{c c c c c c c}
\hline \hline
$\Ek$ & $\Pran$ & $\Ra$ & $\Rey$ & $\lpeak$ & $\lpeak_r(s=0.5)$ & $(N_s,M)$
\\ \hline 
$10^{-6}$ &  $10^{-2}$ & $6.000\times10^{7}$ & $1352 \pm 171$ & $0.21298 \pm 0.01937$ & $0.22627 \pm 0.10272$ & $(600,96)$\\ 
$10^{-6}$ & $10^{-2}$ & $7.500\times10^{7}$ & $1778 \pm 186$ & $0.21450 \pm 0.02458$ & $0.23513 \pm 0.07598$ & $(600,96)$\\ 
$10^{-6}$ &  $10^{-2}$ & $1.050\times10^{8}$ & $2407 \pm 215$ & $0.22954 \pm 0.02506$ & $0.21932 \pm 0.08423$ & $(600,96)$\\ 
$10^{-6}$ &  $10^{-2}$ & $1.650\times10^{8}$ & $3239 \pm 276$ & $0.23947 \pm 0.02915$ & $0.24426 \pm 0.10119$ & $(600,128)$\\ 
$10^{-6}$ &  $10^{-2}$ & $2.850\times10^{8}$ & $4406 \pm 334$ & $0.25416 \pm 0.03188$ & $0.23071 \pm 0.08192$ & $(600,128)$\\
\hline
$10^{-7}$ & $10^{-2}$ & $8.100\times 10^{8}$ & $2588 \pm 363$ & $0.08521 \pm 0.02132$ & $0.12136 \pm   0.03892$ & $(1000,128)$ \\
$10^{-7}$ & $10^{-2}$ & $9.000\times 10^{8}$ & $3015 \pm 208$ & $0.08748 \pm 0.02315$ & $0.12486 \pm   0.03539$ & $(1000,128)$ \\ 
$10^{-7}$ & $10^{-2}$ & $9.900\times 10^{8}$ & $3459 \pm 253$ & $0.08721 \pm 0.02358$ & $0.12366 \pm   0.03594$ & $(1000,128)$ \\
$10^{-7}$ & $10^{-2}$ & $1.050\times 10^{9}$ & $3569 \pm 260$ & $0.09150 \pm 0.02588$ & $0.12645 \pm   0.03666$ & $(1000,128)$ \\ 
$10^{-7}$ & $10^{-2}$ & $1.170\times 10^{9}$ &  $4204 \pm 244$ & $0.11126 \pm 0.01482$ & $0.12652 \pm   0.03993$ & $(1000,128)$ \\ 
$10^{-7}$ & $10^{-2}$ & $1.350\times 10^{9}$ & $4459 \pm 143$ & $0.09127 \pm 0.02666$ & $0.13376 \pm    0.05529$ & $(1000,256)$ \\ 
$10^{-7}$ & $10^{-2}$ & $1.440\times 10^{9}$ & $4989 \pm 361$ & $0.11824 \pm 0.01319$  & $0.13311 \pm   0.04089$ & $(1000,256)$ \\ 
$10^{-7}$ & $10^{-2}$ & $1.500\times 10^{9}$ & $5278 \pm 510$ & $0.09784 \pm 0.03400$ & $0.14291 \pm    0.03899$ & $(1000,256)$ \\ 
$10^{-7}$ & $10^{-2}$ & $1.725\times 10^{9}$ &  $5689 \pm 298$ & $0.11576 \pm 0.01000$ & $0.14365 \pm   0.04867$ & $(1200,384)$ \\ 
$10^{-7}$ & $10^{-2}$ & $2.070\times 10^{9}$ &  $6588 \pm 274$ & $0.12681 \pm 0.01505$  & $0.14959 \pm   0.08607$ & $(1200,384)$\\ 
$10^{-7}$ & $10^{-2}$ & $2.490\times 10^{9}$ &  $7411 \pm 424$ & $0.13164 \pm 0.01194$  & $0.16849 \pm   0.08160$ &  $(1200,384)$ \\ 
$10^{-7}$ & $10^{-2}$ & $3.000\times 10^{9}$ & $8659 \pm 559$ & $0.14461 \pm 0.01330$ & $0.18763 \pm   0.08680$ &$(1200,384)$ \\ 
$10^{-7}$ & $10^{-2}$ & $3.600\times 10^{9}$ & $9214 \pm 456$ & $0.14711 \pm 0.01796$ & $0.19540 \pm   0.07277$ & $(1200,384)$ \\ 
\hline
$10^{-8}$ & $10^{-2}$ & $1.050\times 10^{10}$ & $4678 \pm 223$ & $0.04566 \pm 0.00430$ & $0.04990 \pm  0.00832$ & $(1600,150)$ \\ 
$10^{-8}$ & $10^{-2}$ & $1.275\times 10^{10}$ & $6359 \pm 321$ & $0.04981 \pm 0.00512$ & $0.05569 \pm  0.01003$ & $(1600,150)$ \\ 
$10^{-8}$ & $10^{-2}$ & $1.500\times 10^{10}$ & $7826 \pm 390$ & $0.05236 \pm 0.00553$ & $0.05996 \pm  0.00974$ & $(1600,256)$ \\ 
$10^{-8}$ & $10^{-2}$ & $1.950\times 10^{10}$ & $11030 \pm 454$ & $0.05919 \pm 0.00911$ & $0.06486 \pm  0.00704$ & $(1600,384)$ \\ 
$10^{-8}$ & $10^{-2}$ & $2.250\times 10^{10}$ & $12561 \pm 352$ & $0.06338 \pm 0.01165$ & $0.06635 \pm  0.00832$ & $(1600,384)$ \\ 
\hline
$10^{-8}$ & $10^{-1}$ & $7.500\times 10^{10}$ & $2279 \pm 53$ & $0.03073 \pm 0.00480$ & & $(2000,400)$ \\ 
$10^{-8}$ & $10^{-1}$ & $1.200\times 10^{11}$ & $3709 \pm 68$ & $0.03746 \pm 0.00502$ & &$(2000,400)$ \\ 
$10^{-8}$ & $10^{-1}$ & $3.000\times 10^{11}$ & $8485 \pm 187$ & $0.05506 \pm 0.00807$ & & $(2000,500)$ \\ 
$10^{-8}$ & $10^{-1}$ & $4.500\times 10^{11}$ & $12953 \pm 287$ & $0.06837 \pm 0.01340$ & & $(2000,700)$ \\ 
$10^{-8}$ & $10^{-1}$ & $7.500\times 10^{11}$ & $21623 \pm 421$ & $0.08380 \pm 0.02534$ & & $(2000,1024)$ \\ 
\hline
$10^{-9}$ & $10^{-2}$ & $1.500\times 10^{11}$ & $7458 \pm 271$ & $0.02350 \pm 0.00389$ & $0.02104 \pm   0.00307$ & $(2000,384)$ \\ 
$10^{-9}$ & $10^{-2}$ & $1.875\times 10^{11}$ & $10026 \pm 238$ & $0.02624 \pm 0.00423$ & $0.02344 \pm  0.00306$ & $(2000,384)$ \\ 
$10^{-9}$ & $10^{-2}$ & $2.250\times 10^{11}$ & $12346 \pm 511$ & $0.02809 \pm 0.00387$ & $0.02528 \pm  0.00458$ & $(2000,384)$  \\ 
$10^{-9}$ & $10^{-2}$ & $2.700\times 10^{11}$ & $15300 \pm 428$ & $0.03056 \pm 0.00432$ & $0.02775 \pm  0.00312$ & $(2000,384)$  \\ 
$10^{-9}$ & $10^{-2}$ & $4.500\times 10^{11}$ & $32274 \pm 487$ & $0.04013 \pm 0.00661$ & $0.03805 \pm  0.00443$ & $(2000,500)$  \\ 
$10^{-9}$ & $10^{-2}$ & $9.000\times 10^{11}$ & $56251 \pm 610$ & $0.04994 \pm 0.01229$ & $0.04513 \pm  0.00446$ & $(2000,500)$ \\ 
$10^{-9}$ & $10^{-2}$ & $1.200\times 10^{12}$ & $74423 \pm 2222$ & $0.05674 \pm 0.01101$ & $0.05007 \pm   0.00645$ & $(2000,512)$ \\ 
\hline \hline
\end{tabular}
\end{center}
\caption{List of the input and output parameters for the simulations performed with the QG model. The azimuthal lengthscale $\lpeak$ 
is averaged in radius between $s=0.1$ and $0.6$. The radial lengthscale $\lpeak_r$ is given at radius $s=0.5$. The last column gives the numerical resolution
with $N_s$ the number of grid points in radius and $M$ the truncation order of the Fourier decomposition in azimuth.}
\label{tab:QG}
\end{table}

\setcounter{table}{0}   
\begin{table}
\begin{center}
\begin{tabular}{c c c c c c c}
\hline \hline
$\Ek$ & $\Pran$ & $\Ra$ & $\Rey$ & $\lpeak$ & $\lpeak_r(s=0.5)$ & $(N_s,M)$
\\ \hline 
$10^{-9.5}$ & $10^{-2}$ & $4.575\times 10^{11}$ & $6716 \pm 165$ & $0.01465 \pm 0.00236$ & $0.01164 \pm  0.00108$ & $(2000,512)$ \\ 
$10^{-9.5}$ & $10^{-2}$ & $4.725\times 10^{11}$ & $7259 \pm 205$ & $0.01492 \pm 0.00247$ & $0.01218 \pm  0.00123$ & $(2000,512)$ \\ 
$10^{-9.5}$ & $10^{-2}$ & $4.950\times 10^{11}$ & $7949 \pm 204$ & $0.01531 \pm 0.00263$ & $0.01298 \pm  0.00110$ & $(2000,512)$ \\ 
$10^{-9.5}$ & $10^{-2}$ & $5.445\times 10^{11}$ & $9220 \pm 205$ & $0.01606 \pm 0.00284$ & $0.01385 \pm  0.00129$ & $(2000,512)$ \\ 
$10^{-9.5}$ & $10^{-2}$ & $6.000\times 10^{11}$ & $10379 \pm 221$ & $0.01683 \pm 0.00303$ & $0.01464 \pm  0.00168$ & $(2000,512)$ \\ 
$10^{-9.5}$ & $10^{-2}$ & $6.600\times 10^{11}$ & $11621 \pm 265$ & $0.01774 \pm 0.00329$ & $0.01498 \pm  0.00128$ & $(2000,600)$ \\ 
$10^{-9.5}$ & $10^{-2}$ & $7.350\times 10^{11}$ & $13016 \pm 235$ & $0.01870 \pm 0.00363$ & $0.01594 \pm  0.00156$ & $(2000,600)$  \\ 
$10^{-9.5}$ & $10^{-2}$ & $8.100\times 10^{11}$ & $14528 \pm 288$ & $0.01949 \pm 0.00368$ & $0.01664 \pm  0.00150$ & $(2000,600)$ \\ 
$10^{-9.5}$ & $10^{-2}$ & $9.000\times 10^{11}$ & $16382 \pm 285$ & $0.02036 \pm 0.00359$ & $0.01725 \pm  0.00203$ & $(2000,650)$ \\ 
$10^{-9.5}$ & $10^{-2}$ & $9.900\times 10^{11}$ & $18144 \pm 341$ & $0.02165 \pm 0.00450$ & $0.01778 \pm  0.00153$ & $(2000,650)$ \\ 
$10^{-9.5}$ & $10^{-2}$ & $1.350\times 10^{12}$ & $26383 \pm 745$ & $0.02480 \pm 0.00442$ & $0.02068 \pm  0.00283$ & $(2000,650)$ \\ 
\hline
$10^{-10}$ & $10^{-2}$ & $1.725\times 10^{12}$ & $8481 \pm 129$ & $0.00992 \pm 0.00181$ & $0.00771 \pm 0.00062$ & $(1600,512)$ \\ 
$10^{-10}$ & $10^{-2}$ & $1.800\times 10^{12}$ & $9592 \pm 152$  & $0.01023 \pm 0.00197$ & $0.00841 \pm   0.00075$ & $(1600,512)$ \\ 
$10^{-10}$ & $10^{-2}$ & $1.950\times 10^{12}$ & $11156 \pm 168$ & $0.01073 \pm 0.00214$ & $0.00906 \pm   0.00075$ & $(1600,512)$ \\ 
$10^{-10}$ & $10^{-2}$ & $2.100\times 10^{12}$ & $12421 \pm 136$ & $0.01117 \pm 0.00226$ & $0.00944 \pm   0.00072$ & $(1600,512)$ \\ 
$10^{-10}$ & $10^{-2}$ & $2.400\times 10^{12}$ & $14459 \pm 182$ & $0.01189 \pm 0.00243$ & $0.00999 \pm   0.00097$ & $(1600,512)$ \\ 
$10^{-10}$ & $10^{-2}$ & $2.700\times 10^{12}$ & $15906 \pm 173$ & $0.01261 \pm 0.00269$ & $0.01052 \pm   0.00075$ & $(2000,512)$\\ 
$10^{-10}$ & $10^{-2}$ & $3.150\times 10^{12}$ & $18587 \pm 214$ & $0.01343 \pm 0.00285$ & $0.01103 \pm   0.00126$ & $(2000,512)$ \\ 
$10^{-10}$ & $10^{-2}$ & $3.900\times 10^{12}$ & $22573 \pm 279$ & $0.01488 \pm 0.00326$ & $0.01209 \pm   0.00105$ & $(3000,700)$ \\ 
$10^{-10}$ & $10^{-2}$ & $4.200\times 10^{12}$ & $24263 \pm 324$ & $0.01534 \pm 0.00332$ & $0.01235 \pm  0.00147$ & $(3000,700)$ \\ 
$10^{-10}$ & $10^{-2}$ & $4.500\times 10^{12}$ & $26152 \pm 289$ & $0.01581 \pm 0.00330$ & $0.01259 \pm  0.00142$ & $(3000,700)$ \\ 
$10^{-10}$ & $10^{-2}$ & $5.250\times 10^{12}$ & $31018 \pm 316$ & $0.01729 \pm 0.00395$ & $0.01358 \pm  0.00119$ & $(3000,900)$\\ 
$10^{-10}$ & $10^{-2}$ & $5.700\times 10^{12}$ & $34032 \pm 534$ & $0.01781 \pm 0.00377$ & $0.01374 \pm  0.00106$ & $(3000,900)$\\ 
$10^{-10}$ & $10^{-2}$ & $6.300\times 10^{12}$ & $38381 \pm 672$ & $0.01897 \pm 0.00447$ & $0.01460 \pm  0.00163$ & $(3000,1024)$ \\ 
\hline
$10^{-10.5}$ & $10^{-2}$ & $7.200 \times 10^{12} $ & $12779 \pm 84$ & $0.00709 \pm 0.00161$ & $0.00596 \pm   0.00043$ & $(3500,1024)$ \\ 
$10^{-10.5}$ & $10^{-2}$ & $7.500 \times 10^{12}$ & $13723 \pm 113$ & $0.00726 \pm 0.00167$ & $0.00621 \pm   0.00077$ & $(3500,1024)$ \\ 
$10^{-10.5}$ & $10^{-2}$ & $9.000\times 10^{12}$ & $17621 \pm 142$ & $0.00810 \pm 0.00192$ & $0.00671 \pm   0.00044$ & $(3500,1024)$ \\ 
$10^{-10.5}$ & $10^{-2}$ & $1.050\times 10^{13}$ & $20540 \pm 145$ & $0.00871 \pm 0.00211$ & $0.00717 \pm   0.00045$ & $(3500,1024)$ \\ 
$10^{-10.5}$ & $10^{-2}$ & $1.200\times 10^{13}$ & $23086 \pm 163$ & $0.00924 \pm 0.00228$ & $0.00740 \pm  0.00050$ & $(3500,1024)$ \\ 
$10^{-10.5}$ & $10^{-2}$ & $1.350\times 10^{13}$ & $25577 \pm 202$ & $0.00976 \pm 0.00240$ & $0.00779 \pm   0.00055$ & $(3500,1024)$ \\ 
$10^{-10.5}$ & $10^{-2}$ & $1.500\times 10^{13}$ & $28174 \pm 256$ & $0.01026 \pm 0.00251$ & $0.00795 \pm   0.00057$ & $(3500,1024)$ \\ 
$10^{-10.5}$ & $10^{-2}$ & $2.250\times 10^{13}$ & $41907 \pm 290$ & $0.01221 \pm 0.00285$ & $0.00945 \pm   0.00094$ & $(3500,1024)$ \\ 
\hline
$10^{-11}$ & $10^{-2}$ & $2.700 \times 10^{13}$ & $15800 \pm 83$ & $0.00462 \pm 0.00112$ & $0.00403 \pm   0.00022$ & $(2400,950)$ \\ 
$10^{-11}$ & $10^{-2}$ & $2.850 \times 10^{13}$ & $17897 \pm 91$ & $0.00482 \pm 0.00120$ & $0.00427 \pm   0.00018$ & $(2400,1024)$ \\ 
$10^{-11}$ & $10^{-2}$ & $3.300 \times 10^{13}$ & $21754 \pm 99$ & $0.00536 \pm 0.00137$ & $0.00455 \pm   0.00027$ & $(4000,1100)$ \\ 
$10^{-11}$ & $10^{-2}$ & $3.750 \times 10^{13}$ & $25243 \pm 127$ & $0.00579 \pm 0.00149$ & $0.00485 \pm  0.00016$ & $(4000,1536)$ \\ 
$10^{-11}$ & $10^{-2}$ & $4.200 \times 10^{13}$ & $28140 \pm 108$ & $0.00622 \pm 0.00166$ &$0.00496 \pm    0.00026$ & $(4000,2048)$ \\ 
$10^{-11}$ & $10^{-2}$ & $4.500 \times 10^{13}$ & $29846 \pm 85$ & $0.00639 \pm 0.00168$ & $0.00514 \pm 0.00019$ & $(4000,2048)$ \\ 
$10^{-11}$ & $10^{-2}$ & $5.250 \times 10^{13}$ & $33657 \pm 211$ & $0.00682 \pm 0.00189$ & $0.00537 \pm 0.00025$ & $(4000,2048)$ \\ 
\hline \hline
\end{tabular}
\end{center}
\caption{Continued.}
\end{table}


\newpage
\section*{Extended Data Table~\ref{tab:3D}: Details of the 3D numerical simulations}

\begin{table}[!ht]
\begin{center}
\begin{tabular}{c c c c c c}
\hline \hline
$\Ek$ & $\Pran$ & $\Ra$ & $\Rey$ & $\lpeak$ & $(N_r, L, M)$
\\ \hline 
$10^{-6}$ & $10^{-2}$ & $6.02 \times 10^7$ & $1160$ & $0.26800 \pm 0.02910$ & $(560, 150, 127)$ \\ 
$10^{-6}$ & $10^{-2}$ & $1.00 \times 10^8$ & $2050$ & $0.31500 \pm 0.03700$ & $(560, 150, 127)$ \\ 
$10^{-6}$ & $10^{-2}$ & $2.00 \times 10^8$ & $3370$ & $0.37600 \pm 0.04820$ & $(560, 150, 127)$ \\ 
\hline
$10^{-7}$ & $10^{-2}$ & $7.00 \times 10^8$ & $1470$ & $0.09430 \pm 0.01360$ & $(1152, 199, 159)$ \\ 
$10^{-7}$ & $10^{-2}$ & $1.10 \times 10^9$ & $3340$ & $0.11500 \pm 0.00894$ & $(1152, 199, 159)$ \\ 
$10^{-7}$ & $10^{-2}$ & $1.25 \times 10^9$ & $3800$ & $0.11900 \pm 0.00788$ & $(1152, 199, 159)$ \\ 
\hline
$3 \times 10^{-8}$ & $10^{-2}$ & $3.00 \times 10^9$ & $2897$ & $0.06480 \pm 0.00346$ & $(1568, 277, 255)$ \\ 
$3 \times 10^{-8}$ & $10^{-2}$ & $4.00 \times 10^9$ & $3967$ & $0.07190 \pm 0.00362$ & $(1568, 277, 255)$ \\ 
$3 \times 10^{-8}$ & $10^{-2}$ & $5.10 \times 10^9$ & $4600$ & $0.07540 \pm 0.00510$ & $(1568, 277, 255)$ \\ 
\hline
$10^{-8}$ & $10^{-2}$ & $1.50 \times 10^{10}$ & $4780$ & $0.04920 \pm 0.00563$ & $(2016, 351, 319)$ \\ 
$10^{-8}$ & $10^{-2}$ & $2.00 \times 10^{10}$ & $5550$ & $0.05250 \pm 0.00607$ & $(2016, 351, 319)$ \\ 
$10^{-8}$ & $10^{-2}$ & $2.50 \times 10^{10}$ & $6100$ & $0.05600 \pm 0.00788$ & $(2016, 351, 319)$ \\ 
\hline \hline
\end{tabular}
\end{center}
\caption{List of the input and output parameters for the simulations performed with the 3D model. The azimuthal lengthscale $\lpeak$
is averaged in radius between $s=0.1$ and $0.6$. The last column gives the numerical resolution
with $N_r$ the number of grid points in radius, $L$ and $M$ the truncation degree and order of the spherical harmonics.
Hyperviscosity was used in all these 3D runs, with viscosity depending on spherical harmonic degree $\ell$, but only for $\ell > 0.9L$. We use $\nu(\ell) = \nu_0$ for $\ell < \ell_c = 0.9 L$ and $\nu(\ell) = \nu_0 q^{\ell-\ell_c}$ for $\ell \geq \ell_c$. We set $q = (\nu_{max}/\nu_0)^{1/(L-\ell_c)}$ and $\nu_{max} \leq 100$.}
\label{tab:3D}
\end{table}


\end{document}